\documentclass{PoS}

\title{Alternative dark matter candidates: Axions}

\ShortTitle{Alternative dark matter candidates: Axions}

\author{\speaker{Andreas Ringwald}
\\
        Deutsches Elektronen-Synchrotron DESY\\
        Notkestr. 85 -- D-22607 Hamburg -- Germany\\
        E-mail: \email{andreas.ringwald@desy.de}}

\abstract{
The axion is arguably one of the best motivated candidates for dark matter. 
For a decay constant $\gtrsim 10^9$ GeV, axions are dominantly produced
non-thermally in the early universe and hence are ``cold", their velocity dispersion
being small enough to fit to large scale structure. Moreover, such a large decay
constant ensures the stability at cosmological time scales 
and its behaviour as a collisionless fluid at cosmological length scales. 
Here, we review the state of the art of axion dark matter predictions and 
of experimental efforts to search for axion dark matter in laboratory experiments.
}

\FullConference{Neutrino Oscillation Workshop\\
		4 - 11 September, 2016\\
		Otranto (Lecce, Italy)}

\begin{document}

\section{Introduction}

Theoretical particle physicsists are very imaginative in proposing particle candidates 
for cold dark matter. Arguably the best motivated ones are those which arise in 
high-energy completions of the Standard Model (SM) which solve also other fundamental problems
of particle physics and cosmology, e.g. neutralinos or gravitinos in supersymmetric extensions
of the SM which solve also the hierarchy problem, the lightest sterile neutrino in the $\nu$MSM 
which explains also neutrino masses
and mixing as well as the baryon asymmetry of the Universe via the introduction of 
right-handed sterile neutrinos featuring a Majorana mass,  or the axion in the Peccei-Quinn
extended SM (PQSM) \cite{Peccei:1977hh} which solves also the strong CP problem, namely the puzzle why the theta parameter
in QCD is so small, $|\theta|<10^{-10}$.

\section{Axionic solution of the strong CP problem}

In PQSM models, the SM is extended by a singlet complex scalar field $\sigma$ 
featuring a global $U(1)_{\rm PQ}$ symmetry which is spontaneously broken at 
a symmetry breaking scale $v_{\rm PQ}$ \cite{Kim:1979if,Shifman:1979if,Zhitnitsky:1980tq,Dine:1981rt}. 
Correspondingly, the excitation $\rho$ of the modulus of the PQ field, 
$\sigma (x)=\frac{1}{\sqrt{2}}[v_{\rm PQ}+\rho(x)]\exp(i\frac{A(x)}{v_{\rm PQ}})$, is massive,
while the excitation of the phase $A(x)$, the axion field, is a light pseudo Nambu-Goldstone boson \cite{Weinberg:1977ma,Wilczek:1977pj}. 
The SM quarks or new hypothetic exotic quarks are supposed to carry PQ charges such that
$U(1)_{\rm PQ}$ is anomalously broken due to a gluonic triangle anomaly, 
\begin{equation}
\label{anomaly}
\partial_\mu J_{U(1)_{\rm PQ}}^\mu \supset 
-\frac{\alpha_s}{8\pi}\,N\, G_{\mu\nu}^a \tilde G^{a\,\mu\nu} 
,
\end{equation}
where $N$ is a model-dependent anomaly coefficient, and $G$ and $\tilde G$ 
 is the gluonic field strength tensor and its dual, 
respectively. Correspondingly, the axion field acts as a space-time dependent 
theta angle, $-\pi\leq \theta (x)\equiv A(x)/f_A\equiv N\, A(x)/v_{\rm PQ}\leq \pi$, i.e. 
couples linearly to the gluonic topological charge density,
$\mathcal L \supset 
-\theta (x)\,q(x) \equiv
-\frac{\alpha_s}{8\pi}\,\frac{A(x)}{f_A}\,G_{\mu\nu}^a \tilde G^{a\,\mu\nu} 
$,
and its self-interactions $V(\theta )$ are determined by the theta dependence of the QCD vacuum energy, 
i.e. they can be 
obtained from the QCD vacuum-to-vacuum amplitude 
 in the presence of a $\theta$ term,  $Z(\theta ,0)$, 
via the relation
\begin{equation}
\mathcal L \supset \int d^4x\,V(\theta )=-\ln \left(\frac{Z(\theta,0)}{Z(0,0)}\right)
.
\end{equation} 
Exploiting the fact that $Z(\theta ,0)$ can be written as a Fourier series, 
$Z(\theta ,0)=\sum_{Q=-\infty}^{+\infty}\textrm{e}^{i\theta Q}Z_Q(0)$, of positive partition functions 
$Z_Q(0)$
of fixed topological charge $Q\equiv\int d^4x\, q(x)$, it can be shown that 
$V(\theta )$ 
has an absolute minimum at $\theta =0$ \cite{Vafa:1984xg}. Therefore, the vacuum 
expection value of $\theta(x)$ vanishes and the strong CP problem is solved. 

The axion mass is given by the second derivative of the QCD vacuum energy 
at $\theta =0$ and can thus be expressed in terms of the topological susceptibility,
$m_A^2 f_A^2 \equiv \chi (0)  \equiv \int d^4x\, \langle q(x)\,q(0)\rangle$.
The latest lattice calculation of the latter found $\chi (0)= [75.6(1.8)(0.9) {\rm MeV}]^4$ \cite{Borsanyi:2016ksw}.
Alternatively, the theta dependence of the vacuum energy in QCD can be determined via chiral perturbation theory,
yielding $\chi (0) \simeq m_\pi^2 f_\pi^2\, \frac{m_u m_d}{(m_u+m_d)^2}$ \cite{Weinberg:1977ma}. A recent calculation, including NLO corrections, found 
$\chi (0)= [75.5(5) {\rm MeV}]^4$ \cite{diCortona:2015ldu}, which agrees beautifully with the lattice result and leads to 
the axion mass prediction 
\begin{equation}
\label{zeroTma}
m_A= 
{57.0(7)\,   \left(\frac{10^{11}\rm GeV}{f_A}\right)\mu \textrm{eV}, }
\end{equation} 
in terms of its decay constant $f_A$.

\section{Axion cold dark matter}

\subsection{Evolution of the axion field in the hot primordial plasma}

The evolution of the axion field in the hot primordial plasma is determined by the solutions 
of 
\begin{equation}
\label{KG}
\ddot \theta + 3 H(T(t)) \dot\theta -  \frac{1}{a^2(t)} \nabla \theta = - \frac{1}{f_A^2} \frac{\partial}{\partial\theta}
V(\theta , T(t)),
\end{equation}
where $H=\frac{\dot a}{a}$ is the expansion rate of the Universe in terms of the scale factor $a(t)$ and
$V(\theta,T)$ is the temperature dependent axion potential which equals the free energy density in QCD 
as a funtion of $\theta$ and $T$. The latter can be calculated at temperatures far above the critical temperature
$T_c\simeq 150$\,MeV of the transition between a hadronic phase and a quark-gluon phase, 
exploiting the semiclassical expansion of the Euclidean path integral about minima of the Euclidean action 
with fixed topological charge -- so called finite-temperature instanton or caloron solutions \cite{Harrington:1978ve}. 
In this case, the partition
function can be described by a dilute gas of instantons (DIGA) with $|Q|=1$,
\begin{equation}
Z(\theta , T) \simeq \sum_{n,\bar n=0}^{\infty}
\left( Z_I(T) \textrm{e}^{i\theta}\right)^n \left( Z_I(T) \textrm{e}^{-i\theta}\right)^{\bar n}
=
\exp \left\{ 2\, Z_I(T) \cos\theta \right\},
\end{equation}
leading to a potential of the form 
\begin{equation}
V(\theta , T) = \chi (T) \left( 1 - \cos\theta \right),
\end{equation} 
where the whole temperature dependence resides in the contribution of a single $|Q|=1$ instanton 
to the path integral, $Z_I(T)$ \cite{Pisarski:1980md},
\begin{equation}
\chi (T) \simeq \frac{2\,Z_I(T)}{\int d^4 x} \propto T^{-(7 + n_f/3)},
\end{equation}
with $n_f$ being the number of active quark flavors at temperature $T$, i.e. for which $m_f\lesssim \pi T$. 

\begin{figure}
\centerline{
\includegraphics[width=0.5\textwidth]{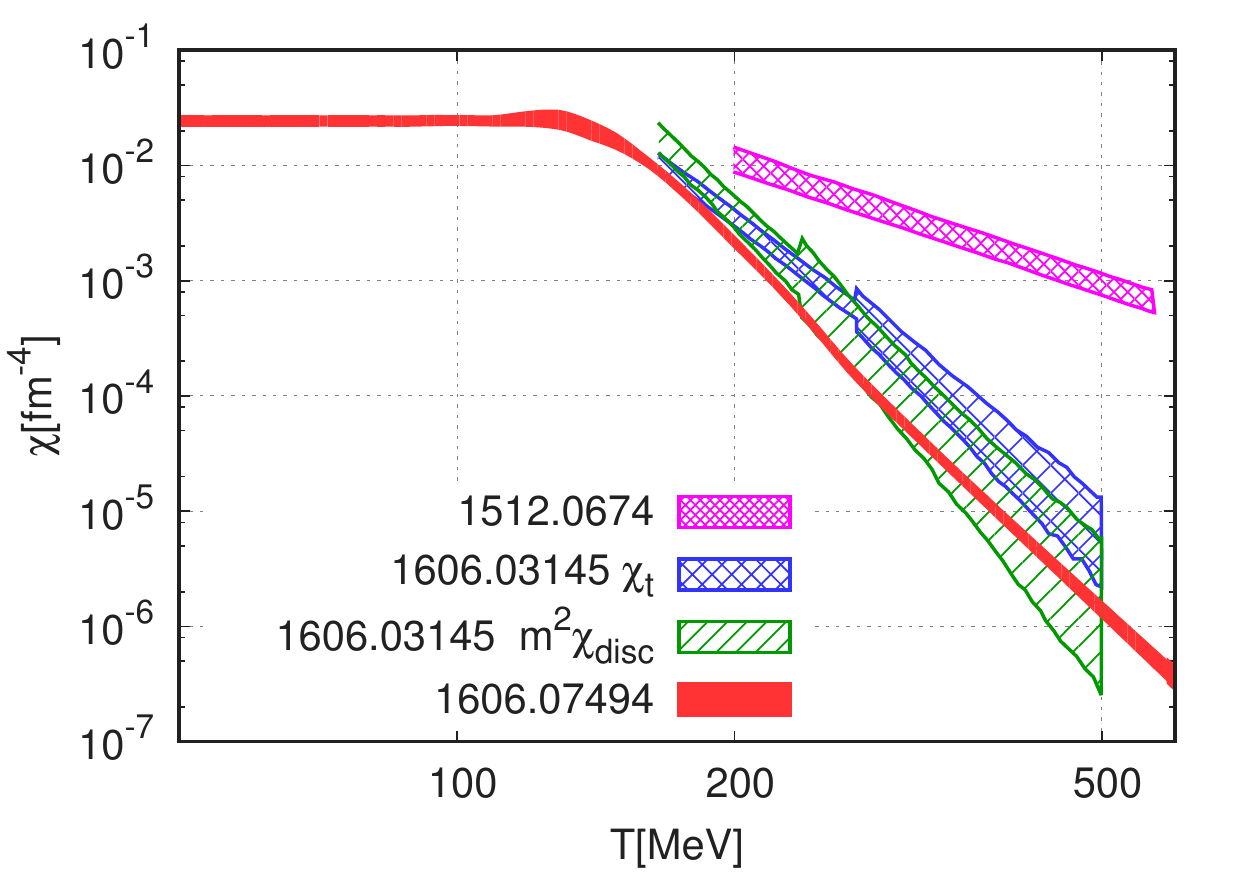}
\includegraphics[width=0.5\textwidth]{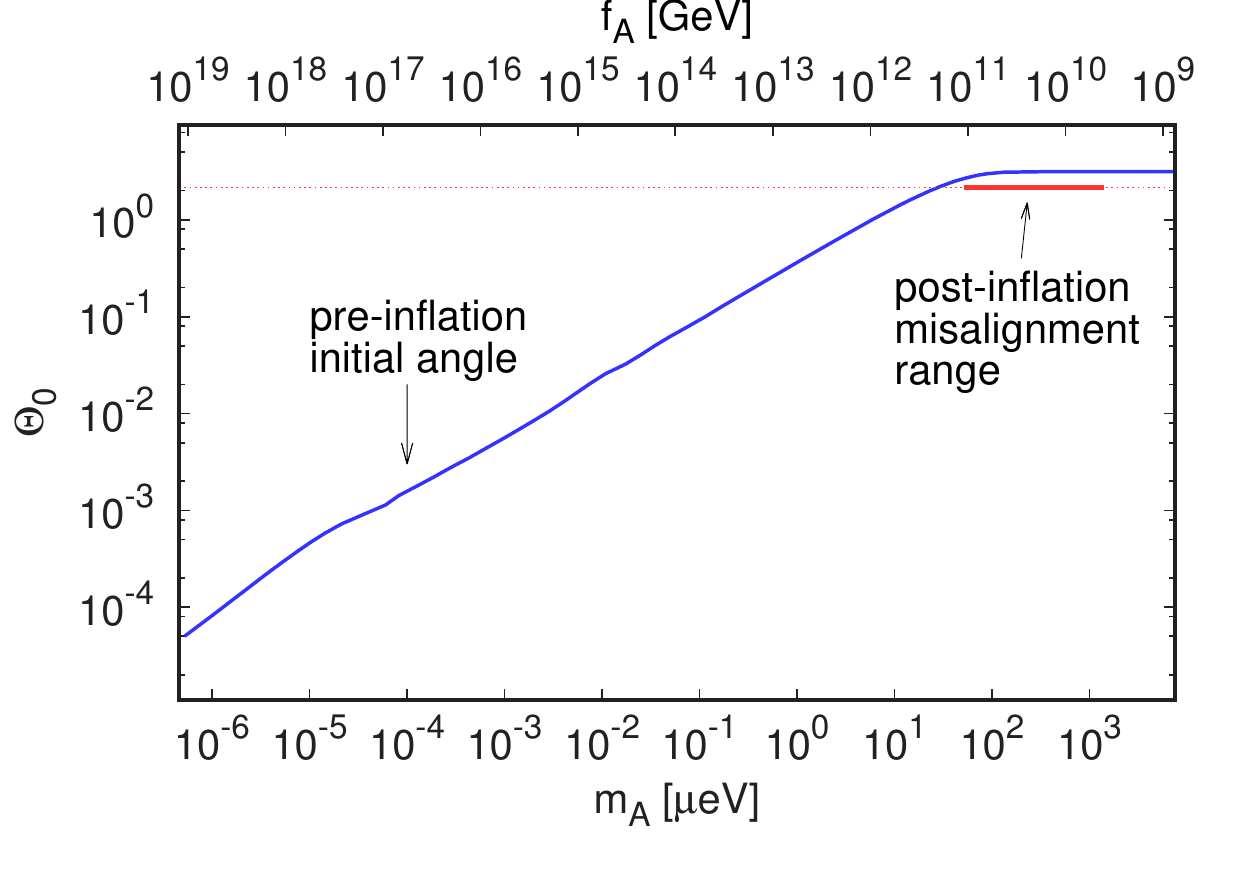}
}
\caption{\label{fig:topsusc}
Left: A compilation of recent lattice results on the topological susceptibility for full QCD as a function of 
temperature \cite{Borsanyi:2016bzg}.
Red band: Borsanyi et al. \cite{Borsanyi:2016ksw}.
Magenta band: Bonati et al. \cite{Bonati:2015vqz}. 
Blue and green bands: Petreczky et al.
\cite{Petreczky:2016vrs}.
Right: Relation between axion mass and initial value 
	$\theta_0$ in order that axions account for all the cold dark matter  
in the Universe \cite{Borsanyi:2016ksw}. 
}
\end{figure}

However, for the prediction of axion dark matter, the temperature range of most interest is around 
a GeV, where the applicability of DIGA is questionable. Fortunately, advances in lattice calculations have allowed
recently to determine the topological susceptibility $\chi(T)$ directly in this temperature region, cf. Fig. \ref{fig:topsusc} (left). 
It is found 
that the temperature slopes are remarkably close to the DIGA prediction. However, DIGA underestimates 
the topological susceptibility by a temperature independent overall  normalization factor of order ten. 
In fact, it seems that an interacting instanton liquid model \cite{Wantz:2009mi} gives a better fit to the normalization 
of the lattice data in the relevant temperature range \cite{Ballesteros:2016xej}. 
In the following, we are basing our axion dark matter estimates on the lattice results for $\chi(T)$ from Ref.  \cite{Borsanyi:2016ksw}.

After PQ symmetry breaking in the very early Universe, the axion field is essentially massless and its
spatial inhomogeneities are erased rapidly over distances of order the size of the causal horizon $H^{-1}$. 
The Hubble friction term in (\ref{KG}) leads to the freezing of the zero momentum mode of the axion
field at its initial value. Later, when the temperature dependent axion mass becomes of order the expansion rate, 
$m_A(T)\equiv \sqrt{\chi(T)}/f_A \simeq 3 H(T)$, which occurs at a temperature of order a GeV, 
the axion field starts to oscillate around the minimum of the effective potential. 
Such a coherent oscillation of the classical axion field corresponds to a coherent state of many, extremely 
non-relativistic axions and thus constitutes cold dark matter \cite{Preskill:1982cy,Abbott:1982af,Dine:1982ah}. 

If the PQ symmetry is broken before and during inflation and not restored afterwards 
({\em pre-inflationary PQ symmetry breaking scenario})  
the axion field takes, at the onset of the coherent oscillations, the same initial value $\theta_0$ 
throughout the observable Universe. The predicted amount of axion dark matter depends then 
both on the initial value $\theta_0$ and on $f_A$. As can be seen from the blue line in Fig. \ref{fig:topsusc} (right), 
in this case the axion mass can range between $10^{-12}$ and $10^{-2}$ eV, for initial values $\theta_0$ between
$10^{-4}$ and one.

On the other hand, if the PQ symmetry is restored after inflation ({\em post-inflationary
PQ symmetry breaking scenario}), our present observable Universe contains many patches which had different
initial values of $\theta$ at the time when the oscillations started, corresponding to a spatially averaged initial misalignment value $\theta_0=2.155$, cf. red line in Fig. \ref{fig:topsusc} (right). In this case, $f_A$ is the only free parameter to determine the dark matter contribution 
from the above described vacuum realignment mechanism. Correspondingly, we can obtain 
 a lower limit on $m_A > 28(2)\,\mu$eV from 
the requirement that the realignment contribution should not exceed the observed dark matter abundance.

In the post-inflationary PQ symmetry breaking scenario, additional contributions to axion dark matter arise from
the decay of topological defects \cite{Davis:1986xc,Lyth:1991bb}: axion strings and axion domain walls. 
To infer those we use the results of recent lattice simulations of the evolution of cosmic string and wall networks
\cite{Kawasaki:2014sqa} (see however  Refs. \cite{Fleury:2015aca,Fleury:2016xrz}), taking into account the latest lattice results on $\chi (T)=m_A^2(T)\,f_A^2$   \cite{Borsanyi:2016ksw}.
For $N=1$ (cf. Eq. (\ref{anomaly})), 
string-wall systems are short lived and the fractional axion dark matter contribution to the energy density in the 
Universe is predicted as \cite{Ballesteros:2016xej}
\begin{equation}
\Omega_{A,\rm{tot}}h^2 \approx 1.6^{+1.0}_{-0.7}\times 10^{-2}\times \left(\frac{f_A}{10^{10}\,\mathrm{GeV}}\right)^{1.165},
\label{omega_a_tot_short}
\end{equation}
resulting in a mass range  $50\,\mu\textrm{eV}\lesssim m_A\lesssim 200\,\mu\textrm{eV}$ for which
the axion could account for all of the dark matter in the Universe. 
For $N>1$, on the other hand, string-wall systems are absolutely stable and eventually overclose the Universe. 
A way out is obtained if the PQ symmetry is explicitely broken e.g. by Planck mass suppressed operators. 
In this case, the axion mass range relevant for dark matter can extend all the way up to 15 meV \cite{Ringwald:2015dsf}.

In summary, in the post-inflationary PQ symmetry breaking scenario, the axion mass range relevant for 
dark matter can range from 50 $\mu$eV to 15 meV, cf. the red line in Fig. \ref{fig:topsusc} (right), whose 
lower (upper) boundary value assumes that the contribution from topological defects amounts to 50\% (99\%) 
of the dark matter in the Universe. 
Remarkably \cite{Ringwald:2015lqa,Giannotti:2015kwo}, an axion with a mass in the meV range may at the same
time explain the accumulating hints of excessive energy losses of stars in various stages of their evolution -- red giants, 
helium burning stars, white dwarfs, and neutron stars.

\section{Axion dark matter experiments}

As we have seen, the mass range relevant for axion dark matter is very wide, cf. Fig. \ref{fig:topsusc} (right).
Fortunately, there are a number of active and proposed axion dark matter direct detection 
experiments covering a sizeable portion of it. Most of these experiments exploit resonance effects in order
to enhance the signal. 

CASPEr \cite{Budker:2013hfa} in Mainz will exploit nuclear magnetic resonance techniques to search for oscillating nuclear electric dipole moments induced by the oscillations of the axion dark matter field. It is expected to be sensitive in the
mass region $10^{-12}\,\textrm{eV}\lesssim m_A\lesssim 10^{-9}\,\textrm{eV}$.  

ABRACADABRA \cite{Kahn:2016aff} at MIT aims at searches exploiting a toroidal magnet acting as an oscillating current
ring (ABRACADABRA), sensitive in the mass region $10^{-9}\,\textrm{eV}\lesssim m_A\lesssim 10^{-6}\,\textrm{eV}$.

The currently running experiments ADMX \cite{Stern:2016bbw} in Seattle and X3 \cite{Brubaker:2016ktl} in Yale, and the  experiment 
CULTASK \cite{Chung:2016ysi} currently in construction in South Korea are searching for 
resonant excitation of modes due to axion
photon conversion in microwave cavities in solenoids, in the mass range 
$10^{-6}\,\textrm{eV}\lesssim m_A\lesssim 10^{-4}\,\textrm{eV}$.

MADMAX \cite{TheMADMAXWorkingGroup:2016hpc} in Munich and ORPHEUS \cite{Rybka:2014cya} in Seattle are proposed to search for axion dark matter induced 
electromagnetic excitations in an open dielectric and Fabry-Perot resonator, respectively, placed in a 
strong magnetic field, with discovery potential in the range $3\times 10^{-5}\,\textrm{eV}\lesssim m_A\lesssim 
3\times 10^{-4}\,\textrm{eV}$.

QUAX \cite{Barbieri:2016vwg} in Italy searches for electron spin precession induced by the galactic axion wind. 

Nearly all these experiments rely on the mildly model-dependent coupling of the axion to electromagnetism. 
Notable exceptions are CASPEr which exploits the model-independent feature that axion dark matter corresponds to 
an oscillating theta parameter and thus induces an oscillation of nuclear electric dipole moments, and 
QUAX which exploits the strongly model-dependent coupling to the electron spin. In the mass range around an meV, 
these experiments are complemented by ARIADNE \cite{Arvanitaki:2014dfa} in South Korea, searching for fifth forces due to axions, and IAXO \cite{Armengaud:2014gea}, a proposed helioscope searching for solar axions.

In conclusion, 
 there is lots of search space for axion cold dark matter. There are many ongoing projects and new ideas to search directly
for axion cold dark matter in the laboratory.  Stay tuned!

\end{document}